\documentclass[amsmath, amssymb, aps, physrev, 10pt]{revtex4-2}

\usepackage{graphicx}
\usepackage{dcolumn}
\usepackage{bm}
\usepackage{dsfont}
\usepackage{amsmath}
\usepackage{color}
\usepackage{MnSymbol}
\usepackage{subfigure}
\usepackage{ragged2e}
\usepackage{float}
\usepackage{booktabs}
\usepackage{silence}
\usepackage{tikz}
\usetikzlibrary{decorations.pathreplacing}
\usepackage{lineno}

\def\d{\text{d}}
\def\M11{$\mathbb M^{1+1}$}
\def\orderof#1{\mathcal{O}(#1)}

\def\precstar{\prec\hspace{-1.4mm}\star\hspace{0.8mm}}
\def\CCC{\mathbf{C}}
\def\LLL{\mathbf{L}}

\def\cV{\mathcal V}
\def\cI{\mathcal I}
\def\cR{\mathcal R}

\begin{document}

\title{\textbf{Effective curvature coupling of link-based causal set propagators in $1+1$ dimensions}}
\author{Arsim Kastrati}
\email{Contact author: arsim.kastrati@uni-wuerzburg.de}
\affiliation{Faculty for Physics and Astronomy, Julius Maximilians University Würzburg, Am Hubland, 97074 Würzburg, Germany}
\date{\today}


\begin{abstract}
We study how a recently introduced causal-set propagator defined in terms of links responds to curvature. On sprinklings embedded in a flat 1+1-dimensional Minkowski space, this propagator is known to reproduce the massless retarded continuum Greens function on large scales. In conformally flat embeddings with constant curvature $\mathcal R$, the causal order is unchanged, but the link relationship is sensitive to the physical volume of the spanned causal diamond. We show that this volume dependence generates a leading curvature correction to the propagator. On large distances, this correction is equivalent to a continuum scalar propagator with an effective curvature coupling  of the form $\xi \mathcal R$ with the coupling constant $\xi_{\rm eff}=1/6$, equivalently corresponding to an effective mass-squared parameter $m_{\rm eff}^2=\xi_{\rm eff}\cR$. This coupling is not inserted by hand but emerges from the link-based path sum itself, reflecting the microscopic structure of the causal set. Numerical simulations on sprinklings embedded in $\mathrm{AdS}_{1+1}$ and $\mathrm{dS}_{1+1}$ support the predicted curvature response and
indicate that the effective coupling persists as the sprinkling density is
increased. 
\end{abstract}

\maketitle
\parskip 2mm

\section{Introduction}

Causal set theory is an approach to quantum gravity in which spacetime is fundamentally discrete and where all information about the geometry is encoded in the causal order~\cite{surya2025causal}. A causal set $\mathcal C$ is a partially ordered locally finite set of events. Its order relation $\prec$ encodes causal precedence, and the number of elements in a region represents the spacetime volume~\cite{Bombelli1987,Sorkin2003,Surya2019}. It has been shown that, under suitable causality conditions, causal order determines the conformal structure of a Lorentzian metric, while the volume element fixes the remaining conformal factor~\cite{Malament1977,HawkingKingMcCarthy1976}. Thus, the continuum metric is replaced by two discrete ingredients: order and counting.

A basic test of this framework is whether field propagators on a causal set sprinkled into a given spacetime reproduce the corresponding continuum Green's functions. Scalar fields are especially useful because they various aspects of interest: causality, mass, curvature coupling, and boundary conditions~\cite{BirrellDavies1982,Wald1994}. Scalar fields on causal sets can be constructed in several ways. One important route uses discrete d'Alembertian operators, where one constructs an order-theoretic analogue of the wave operator and studies its continuum limit~\cite{BenincasaDowker2010,DowkerGlaser2013,AslanbeigiSaravaniSorkin2014}. Another route, closer to a path-integral picture, is provided by Johnston's hop-and-stop model, where the propagator is written as a weighted sum over causal trajectories~\cite{Johnston2008}. In flat spacetime, this gives the correct retarded Green's functions for suitable hop and stop amplitudes, and related constructions have been studied in curved spacetimes as well~\cite{XDowkerSurya2017}. In this context, the causal matrix
\begin{equation}
\label{Cdef}
\CCC_{ij}
=
\delta(c_j\prec c_i)
=
\begin{cases}
1 & \text{if }c_j\prec c_i,\\
0 & \text{otherwise.}
\end{cases}
\end{equation}
plays a central role because it records causal accessibility. For example, it has been suggested that the free massless propagator of a scalar field on a causal set is essentially given by $\CCC$. However, it is important to note that $\CCC$ contains the full transitive closure of the order, meaning that $\CCC_{ij}=1$ even there are many intermediate events between $c_j$ and $c_i$. Interpreted in terms of paths this means that an arbitrarily distant causal relation is treated as a possible elementary jump. This is useful in some settings, especially in 1+1 dimensions, where the massless retarded propagator is known to be constant within the light cone in the continuum limit. 

However, if propagation is to be built from elementary causal steps, the more primitive objects are the so-called \textit{links}. Two events $c_j,c_i\in\mathcal C$ are said to be linked, denoted as $c_j\precstar c_i$, if there is no other element $c_k$ with $c_j\prec c_k\prec c_i$. The corresponding link matrix $\LLL$ 
\begin{equation}
\label{Ldef}
\LLL_{ij}
\;=\;
\delta(c_j\precstar c_i)
\;=\;
\begin{cases}
1 & \text{if }\CCC_{ij}=1\text{ and }(\CCC^2)_{ij}=0,\\
0 & \text{otherwise.}
\end{cases}
\end{equation}
represents the transitive reduction of the order, retaining only irreducible causal relations. Thus, $\LLL$ can be understood as the adjacency matrix of the causal set. Its power  $(\LLL^n)_{ij}$ counts the number of paths from $c_j$ to $c_i$ consisting of exactly $n$ links.

In Ref.~\cite{hinrichsen2026linkbasedcausalsetpropagators} we took the link matrix $\LLL$ as the central object from which the propagator is built. For causal sets embedded in a $1+1$-dimensional Minkowski spacetime \M11, we found the exponential series $e^\LLL$, averaged over many sprinklings, becomes asymptotically constant and tends to $e^{-\gamma_E}$ at large timelike separation. Normalizing by the factor $\frac12 e^{\gamma_E}$, this reproduces the massless retarded continuum Green's function, which equals $1/2$ inside the future light cone in $1+1$ dimensions. The present paper asks how the Green's function responds to curvature.

To address this question, we consider embeddings with constant curvature. Anti--de Sitter (AdS) spacetime and de Sitter (dS) spacetime are maximally symmetric, have analytically known continuum propagators, and isolate the effect of a nonzero scalar curvature without introducing curvature gradients. They are also important physical testing grounds: de Sitter spacetime appears in cosmology, while anti--de Sitter spacetime plays a central role in the study of fields with timelike boundaries and in the AdS/CFT correspondence~\cite{AvisIshamStorey1978,BreitenlohnerFreedman1982,Maldacena1998,Witten1998,Aharony2000}. The test is especially sharp in 1+1 dimensions, where both AdS and dS can be described in terms of conformal coordinates with the line element
\begin{equation}
\label{eq:conformal-metric-intro}
ds^2=\Omega^2(t,x)(-dt^2+dx^2)\,.
\end{equation}
Consequently, the light cone structure is preserved, while the curvature is encoded solely in the volume of the events. In $1+1$ dimensions, the minimally coupled massless scalar equation is conformally invariant. Consequently, for the continuum reference theory considered here, the retarded Green function remains equal to $1/2$ in the interior of the future light cone. This makes the effect of the link construction particularly transparent: the causal order is unchanged by the conformal factor, whereas link probabilities remain sensitive to the physical volumes of Alexandrov intervals.

Because of the conformal structure, the causal matrix $\CCC$ is insensitive to curvature at the level of causal order. For the link matrix $\LLL$ the situation is different. A link between two events $a$ and $b$ is defined as a causal connection with an \textit{empty} causal diamond (Alexandrov interval) between. For a Poisson sprinkling of density $\rho$, the expectation value of the matrix elements of $\LLL$ is therefore given by
\begin{equation}
\label{eq:link-emptiness-intro}
\langle \LLL_{ab}\rangle=e^{-\rho V(a,b)}\,,
\qquad
V(a,b)=\int_{I(a,b)}\d V\,,
\end{equation}
where $dV=\Omega^2(t,x)\d t \d x$. This means that links are sensitive to the volume of the diamond and thus feel the curvature of the embedding space. 

In this work, we analyze the link-exponential propagator $G \propto \langle e^\LLL\rangle$ on causal sets sprinkled into 1+1-dimensional constant-curvature spacetimes. Our main analytical result is that $G$ is not asymptotically constant in the future light cone. Instead, in leading order of the curvature $\mathcal R$, it picks up a correction of the form
\begin{equation}
\label{eq:main-result-intro}
G_{\rm norm}(a,b)
\simeq
\frac{1}{2}
-
\frac{1}{24}
\int_{I(a,b)}
\mathcal R\,\d V
+\cdots .
\end{equation}
For constant curvature, this correction turns out to be equivalent to a curvature coupling $\xi \mathcal R$ with the effective coupling constant $\xi_{\rm eff} = 1/6$, corresponding to the effective curvature-induced mass parameter
\begin{equation}
\label{eq:xi-induced-intro}
m_{\rm eff}^2
\;=\;
\xi_{\rm eff}\mathcal R
\;=\;
\frac{\mathcal R}{6}.
\end{equation}
Note that this coupling term is not inserted by hand. It emerges as a property of the causal structure and is induced by the volume dependence of the link relation. 

In the following sections, we derive Eq.~\eqref{eq:main-result-intro}-\eqref{eq:xi-induced-intro} perturbatively. If $\tau$ denotes the temporal geodesic distance, the calculations are carried out in the regime
\begin{equation}
\label{eq:regime-intro}
\frac{1}{\rho} 
\;\ll \;
\tau^2
\;\ll \;
\frac{1}{|\mathcal R|}
\,.
\end{equation}
This means that $\tau$ is large compared to the discreteness scale of the causal set, but still small enough so that the curvature effects can be treated perturbatively. After determining the coefficient \(m_{\rm eff}^2=\mathcal R/6\), we compare the numerical data with the corresponding AdS and dS propagators in the continuum. The numerical results suggest that the obtained results are valid even beyond perturbation theory.

An important feature of this result is that the leading curvature coefficient is independent of
the sprinkling density~$\rho$. It controls the approach to the asymptotic regime
$\rho\tau^2\gg1$, but does not appear in the resulting coefficient
$\xi_{\mathrm{eff}}=1/6$. The numerical density study supports this analytical result: increasing
$\rho$ primarily modifies the short-distance region, while the large-distance curvature response
approaches the same density-independent behavior.

Taken together, the analytical and numerical results indicate that the normalized
link-exponential kernel is naturally associated with the retarded Green function of the
effective operator $\Box_g-\mathcal R/6$, rather than with the minimally coupled massless
operator $\Box_g$. While this identification is derived here analytically only to leading order in
curvature, the AdS and dS simulations support it over the full constant-curvature regime
explored numerically.

Interestingly, curvature-coupling terms have also been obtained in studies based on causal-set d'Alembertian operators~\cite{BenincasaDowker2010,DowkerGlaser2013,AslanbeigiSaravaniSorkin2014}. with coefficients that differ from the value $1/6$ found here. This demonstrates that the effective curvature response is not universal across all causal-set discretizations. Whether particular classes of constructions, such as propagators built directly from links, exhibit a characteristic curvature response remains an open question. 

The paper is organized as follows. Section~\ref{sec:flat-reference} recalls link-based propagators suggested in~\cite{hinrichsen2026linkbasedcausalsetpropagators}. Section~\ref{sec:link-propagation-curved} computes the first curvature correction to the one-link and two-link contributions in constant-curvature backgrounds. Section~\ref{sec:curvature-correction} explains how this curvature coefficient propagates through the link-exponential generating function. Finally, section~\ref{sec:constant-curvature-numerics} considers AdS and dS spacetime and compares with numerical simulations. The technical details are given in Appendix~\ref{AppendixCalculation}.

\section{Link paths and the flat reference kernel}
\label{sec:flat-reference}

We first recall the construction of a link-base propagator in a flat embedding space suggested in \cite{hinrichsen2026linkbasedcausalsetpropagators}. For two causally related elements $a\prec b$ with embedding coordinates $t_a,x_a$ and $t_b,x_b$, define
\begin{equation}
\label{eq:Fn-def}
F_n(a,b):=\Bigl\langle(\LLL^n)_{ab}\Bigr\rangle 
\end{equation}
averaged over many sprinklings with density $\rho$. This quantity describes the expected number of $n$-link paths from $a$ to $b$ and must not be confused with a probability distribution. Since powers of the link matrix concatenate paths, integration over an intermediate point $z$ gives the recursion relation
\begin{equation}
\label{eq:recursion-general}
F_{k+l}(a,b)
=
\rho\int_{z \in I(a,b)}\d V\,F_k(a,z)F_l(z,b).
\end{equation}
In a flat $1+1$-dimensional Minkowski spacetime \M11, statistical Lorentz invariance of the sprinkling implies that $F_n$ depends only on the proper time distance
\begin{equation}
\tau\;=\;d(a,b)\;=\;\sqrt{(t_b-t_a)^2-(x_b-x_a)^2}.
\end{equation}
i.e., $F_k(\tau)=\langle F_k(a,b)\rangle_{d(a,b)=\tau}$. Since the spanned causal diamond has the volume $V(\tau)=\tau^2/2$, the one-link contribution is
\begin{equation}
\label{eq:F1-flat}
F_1^{(0)}(\tau)
\;=\;
e^{-\rho V(\tau)}
\;=\;
e^{-\rho\tau^2/2}\,,
\end{equation}
where the superscript $(0)$ indicates that the embedding space is flat. This function is exponentially short-ranged in $\tau$. The first long-range contribution appears in the two-link term $F_2$. In \M11 its exact expression involves the hyperbolic sine integral, but the only part needed here is its asymptotic behavior
\begin{equation}
\label{eq:F2-flat-tail}
F_2^{(0)}(\tau)\simeq \frac{4}{\rho\tau^2},
\qquad
\rho\tau^2\gg1.
\end{equation}
The link-exponential propagator suggested in~\cite{hinrichsen2026linkbasedcausalsetpropagators} is defined by
\begin{equation}
\label{eq:G-def}
G(a,b)\;:=\;
\left\langle(e^{\LLL})_{ab}\right\rangle
\;=\;
\sum_{n=1}^{\infty}\frac{1}{n!}F_n(a,b).
\end{equation}
Note that no independent hop or stop amplitudes are introduced; the link matrix itself defines the propagator. To analyze the large-distance behavior, one introduces an auxiliary parameter $\lambda$:
\begin{equation}
\label{eq:Glambda-def}
G_\lambda(a,b)\;:=\;\left\langle(e^{\lambda\LLL})_{ab}\right\rangle
\;=\;
\sum_{n=1}^{\infty}\frac{\lambda^n}{n!}F_n(a,b),
\qquad
G(a,b)=G_1(a,b).
\end{equation}
Differentiating the series and using the recursion in (\ref{eq:recursion-general}) gives
\begin{equation}
\label{eq:flow-general}
\partial_\lambda G_\lambda(a,b)
\;=\;
F_1(a,b)
+
\rho\int_{I(a,b)}\d V\,F_1(a,z)G_\lambda(z,b).
\end{equation}
For $\rho\tau^2\gg1$, the first term is exponentially small, so the asymptotic behavior is controlled by the convolution integral. To solve this integral equation in \M11, one uses the ansatz
\begin{equation}
\label{eq:flat-ansatz}
G_\lambda^{(0)}(\tau) \;\simeq\; \alpha(\lambda)\tau^{-\beta(\lambda)}.
\end{equation}
For small values of $\lambda$, the functions $\alpha(\lambda)$ and $\beta(\lambda)$ are determined by the algebraic tail of the two-link contribution
\begin{equation}
\label{eq:flat-small-lambda}
G_\lambda^{(0)}(\tau) \;\simeq\; \frac{2\lambda^2}{\rho\tau^2},
\qquad
\lambda\to0^+.
\end{equation}
On the other hand, as shown in Appendix~\ref{subsec:appendix-flat}, the convolution product on the r.h.s. of Eq.~(\ref{eq:flow-general}) can be approximated as
\begin{equation}
\label{eq:flat-conv-result-main}
\rho\int_{I(a,b)}\d V\,F_1^{(0)}(a,z)G^{(0)}_\lambda(z,b)
\;\simeq\;
G_\lambda^{(0)}(\tau)(2\ln\tau+C_\lambda),
\end{equation}
where $C_\lambda$ is independent of $\tau$. Matching this expression with $\partial_\lambda G_\lambda^{(0)}$ gives
\begin{equation}
\label{eq:flat-alpha-beta}
\beta(\lambda)=2-2\lambda,
\qquad
\alpha(\lambda)=
\frac{e^{-\lambda\gamma_E}(\rho/2)^{\lambda-1}}{\Gamma^2(\lambda)}\,,
\end{equation}
implying that
\begin{equation}
\label{eq:flat-limit}
G^{(0)}(\tau)\;=\;G_1^{(0)}(\tau)\;\longrightarrow\; e^{-\gamma_E},
\qquad
\rho\tau^2\gg1.
\end{equation}
Finally, the normalization is adjusted in order to reproduce the known constant value $1/2$ of the continuum retarded propagator in the interior of the future light cone:
\begin{equation}
\label{eq:Gnorm-def}
G_{\rm norm}(\tau)\;:=\;\frac{1}{2}e^{\gamma_E}G(\tau)\,,
\end{equation}
which reproduces the massless retarded continuum Green's function inside the future light cone. This normalization will be used in the following Section.

\section{Curvature correction to the link-path distribution}
\label{sec:link-propagation-curved}
%
So far, we have summarized the construction of the link-based propagator in \M11. We now compute how the functions $F_1$ and $F_2$ change when the sprinkling is embedded in a $1+1$-dimensional spacetime with constant curvature. The goal is to determine the curvature correction in the algebraic tail to leading order. To this end, we perform an expansion in the regime \eqref{eq:regime-intro}. Note that in curved spacetime, the recursion relation \eqref{eq:recursion-general} remains valid, but the volume element and the link probabilities now depend on the curvature.


\subsection{Conformal coordinates and small interval volume}

In conformal coordinates with the line element~\eqref{eq:conformal-metric-intro}, let us introduce null coordinates $u=t+x$ and $v=t-x$. In these coordinates, the line element and the volume element read
\begin{equation}
 ds^2=-\Omega^2(u,v)\d u\,\d v,
\qquad
dV=\frac12\Omega^2(u,v)\d u\,\d v .
\end{equation}
Since the null directions (light cones boundaries) are unchanged, causal precedence can be expressed as
\begin{equation}
x\prec y
\quad\Longleftrightarrow\quad
u_x<u_y \quad \wedge \quad
v_x<v_y.
\end{equation}
Thus the shape of a causal diamond is the same as in flat spacetime; curvature enters only through the volume element. More specifically, if the diamond is spanned by two points $a$ and $b$ with a small proper-time distance $\tau$, its volume is given by (see Appendix~\ref{subsec:appendix-small-diamond})
\begin{equation}
\label{eq:small-volume-main}
V(a,b)\;=\;\frac{\tau^2}{2}-\frac{\cR\tau^4}{96}+\orderof{\tau^6} .
\end{equation}
As can be seen, a positive (negative) curvature decreases (increases) the volume compared to the flat case.

For the following, it will be convenient to introduce the dimensionless interval 
curvature
\begin{equation}
\label{eq:J-intro}
\mathcal V(a,b)
\;:=\;
\int_{I(a,b)}
\mathcal R\,\d V 
\;=\;
\frac{\cR\tau^2}{2}+\orderof{\tau^4}.
\end{equation}

\subsection{One- and two-link contributions $F_1$ and $F_2$}
\begin{figure}[t]
\centering
\includegraphics[width=60mm]{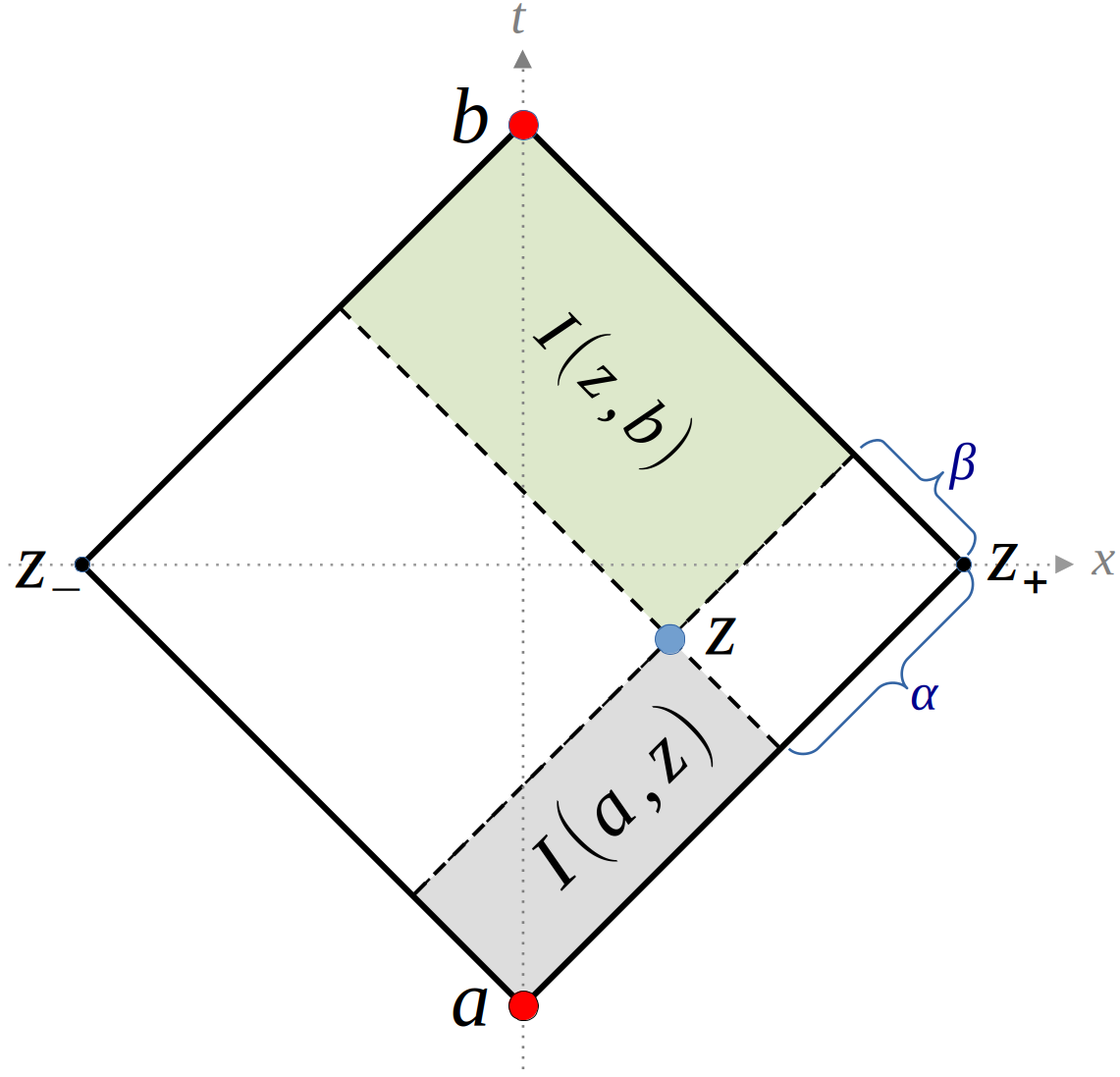}
\caption{
Causal diamond spanned by two points $a$ and $b$. Translation and Lorentz invariance allows us to choose $x_a=x_b=0$. The large diamond $I(a,b)$ has the form of a tilted square. In the recursion relation~(\ref{eq:recursion-general}), one integrates over $z \in I(a,b)$. If the involved functions are short-ranged, the integral is dominated by contributions where $z$ is close to the edges $z_\pm$, the so-called null corners. In this case, the two subintervals $I(a,z)$ and $I(z,b)$ become thin. The variables $\alpha$ and $\beta$ measure the distance from the respective corner along the two null directions.
}
\label{fig:null-corner-region}
\end{figure}

%
On a causal set embedded in a spacetime with constant curvature, the one-link contribution $F_1$ is given by the probability that the causal diamond is empty, that is,
\begin{equation}
\label{eq:F1-curved-main}
F_1(a,b)\;=\;
\langle\LLL_{ab}\rangle
\;=\;e^{-\rho V(a,b)}
\;=\; F_1^{(0)}(\tau)
\left[1+\frac{\rho\cR\tau^4}{96}+\cdots\right],
\end{equation}
where $F_1^{(0)}$ is the corresponding function in flat spacetime given in~\eqref{eq:F1-flat}. As before, this contribution is exponentially short-ranged. The first long-range contribution occurs in the two-link term
\begin{equation}
\label{eq:F2-def-main}
F_2(a,b)
\;=\;
\rho\int_{I(a,b)}\d V\,e^{-\rho V(a,z)}e^{-\rho V(z,b)} .
\end{equation}
As illustrated in Fig.~\ref{fig:null-corner-region}, the two exponential functions localize the integral near the two corners of $I(a,b)$ so that the two strips $I(a,z)$ and $I(z,b)$ become small. This is the same mechanism that produces the algebraic tail in the flat case but now incorporates curvature corrections. 

Suppose that $z$ is close to the right corner $z_+$, as illustrated in the figure. Writing $U=\ell-\alpha$ and $V=-\ell+\beta$ with $\ell=\tau/2$ and $\alpha,\beta\ge0$, the volumes of the subintervals are to leading order linear in $\alpha$ and $\beta$:
\begin{equation}
V(a,z)\simeq A_+\beta,
\qquad
V(z,b)\simeq B_+\alpha\,.
\end{equation}
As shown in Appendix~\ref{subsec:appendix-small-diamond}, including both
null corners gives
\begin{equation}
\label{eq:F2-curvature-main}
F_2(a,b)=
F_2^{(0)}(\tau)
\left[1-\frac{\cR\tau^2}{24}+\cdots\right],
\qquad
F_2^{(0)}(\tau)=\frac{4}{\rho\tau^2}.
\end{equation}
Thus curvature does not remove the algebraic tail found in flat space; it
changes its amplitude. Using the curvature dependent interval volume \eqref{eq:J-intro} the same result becomes
\begin{equation}
\label{eq:F2-J-main}
F_2(a,b)=
F_2^{(0)}(\tau)
\left[1-\frac{1}{12}\cV(a,b)+\cdots\right].
\end{equation}

Equation~\eqref{eq:F2-J-main} gives only the first long-distance term in
the link-path expansion. The full link-exponential kernel contains paths
of all link lengths, so the coefficient \(1/12\) need not remain unchanged
after the exponential resummation. To test this, we now introduce a
\(\lambda\)-dependent curvature-response coefficient, in direct analogy
with the \(\lambda\)-dependent flat amplitude and exponent used in the
flat calculation.

\section{Curvature coefficient of the link-exponential kernel}
\label{sec:curvature-correction}

The flat analysis used the one-parameter generating function
\(G_\lambda\) to switch on paths of increasing link length and then studied
the corresponding flow in \(\lambda\). We follow the same strategy here.
For large timelike separation and to first order in curvature, we write
\begin{equation}
\label{eq:curved-ansatz-main}
G_\lambda(a,b)
\simeq
G_\lambda^{(0)}(\tau)
\left[
1-\zeta(\lambda)\cV(a,b)+\cdots
\right],
\end{equation}
where \(G_\lambda^{(0)}\) is the flat asymptotic function in
Eq.~\eqref{eq:flat-ansatz}. The function \(\zeta(\lambda)\) is not assumed
to be known. It measures how strongly the large-distance kernel responds
to the integrated curvature after paths have been summed up to the
parameter value \(\lambda\).

Its initial value is fixed by the small-\(\lambda\) expansion. Since
\begin{equation}
G_\lambda(a,b)
=
\lambda F_1(a,b)
+
\frac{\lambda^2}{2}F_2(a,b)
+
\orderof{\lambda^3},
\end{equation}
and \(F_1\) is short-ranged for \(\rho\tau^2\gg1\), the first
algebraic tail comes from \(F_2\). Using Eq.~\eqref{eq:F2-J-main}, one
obtains
\begin{equation}
\label{eq:small-lambda-curved}
G_\lambda(a,b)
\simeq
\frac{2\lambda^2}{\rho\tau^2}
\left[
1-\frac{1}{12}\cV(a,b)+\cdots
\right],
\qquad
\lambda\to0^+ .
\end{equation}
Comparison with the ansatz~\eqref{eq:curved-ansatz-main} gives
\begin{equation}
\label{eq:zeta-initial-main}
\zeta(0^+)=\frac{1}{12}.
\end{equation}
The remaining task is to determine whether this initial curvature-response
coefficient is changed by the \(\lambda\)-flow. The main text gives the
structure of the matching, while the radial kernel expansion and the
resulting differential equation for \(\zeta(\lambda)\) are given in
Appendix~\ref{subsec:appendix-curv-flow}.

To see what must be matched, write the convolution term in Eq.~\eqref{eq:flow-general} in radial variables
\begin{equation}
\label{eq:radial-vars-main}
\tau=d(a,b),
\qquad
{\tau'}=d(a,z),
\qquad
{\tau''}=d(z,b).
\end{equation}
Here $T$ is the external separation, $q$ is the short first-link distance, and $r$ is the remaining internal separation. Schematically,
\begin{equation}
\label{eq:schematic-conv-main}
(F_1\circledast G_\lambda)(\tau)
=
\int \d {\tau'}\,\d {\tau''}\,K(\tau;{\tau'},{\tau''})F_1({\tau'})G_\lambda({\tau''}),
\end{equation}
where $K(\tau;{\tau'},{\tau''})$ is the Jacobian converting the integral over $z$ into an integral over the two distances ${\tau'}$ and ${\tau''}$.

The curvature expansion of the integrand has three sources:
\begin{equation}
K=K_0(1+\delta K),
\qquad
F_1=F_1^{(0)}(1+\delta F_1),
\qquad
G_\lambda({\tau''})=G_\lambda^{(0)}({\tau''})[1-\zeta\cV({\tau''})].
\end{equation}
The only subtle bookkeeping point is that the external ansatz already contains the factor $1-\zeta\cV(\tau)$. Therefore we add and subtract this same external factor inside the internal propagator,
\begin{equation}
\label{eq:smart-zero-main}
1-\zeta\cV({\tau''})
=
[1-\zeta\cV(\tau)]
+
\zeta[\cV(\tau)-\cV({\tau''})].
\end{equation}
The first term produces the flat convolution dressed by the external curvature factor. The second term is a genuine relative curvature source.

To first order in curvature, the right-hand side therefore separates as
\begin{equation}
\label{eq:rhs-split-main}
(F_1\circledast G_\lambda)(\tau)
=
[1-\zeta\cV(\tau)](F_1^{(0)}\circledast G_\lambda^{(0)})(\tau)
+
H_{\rm curv}(\tau),
\end{equation}
where
\begin{equation}
\label{eq:Hcurv-def-main}
H_{\rm curv}(\tau)
=
\int \d{\tau''}\,\d{\tau'}\,K_0(\tau;{\tau'},{\tau''})F_1^{(0)}({\tau'})G_\lambda^{(0)}({\tau''})\,\cI(\tau;{\tau'},{\tau''}).
\end{equation}
The relative insertion $\cI$ collects precisely the three first-curvature sources $\delta K$, $\delta F_1$, and $\zeta[\cV(\tau)-\cV({\tau''})]$. For constant curvature $\cR$, these sources are
\begin{equation}
\label{eq:I-main-short}
\cI(\tau;{\tau'},{\tau''})
=
-
\frac{\cR}{24}(\tau^2-{\tau'}^2-{\tau''}^2)
+
\frac{\rho\cR {\tau''}^4}{96}
+
\frac{\zeta\cR }{2}(\tau^2-{\tau''}^2).
\end{equation}
Here the first term comes from the radial kernel, the second from the short one-link factor, and the third from the smart-zero term in Eq.~\eqref{eq:smart-zero-main}. The derivation of these three terms is given in Appendix~\ref{subsec:appendix-kernel}.

The left-hand side of the flow equation is obtained by differentiating Eq.~\eqref{eq:curved-ansatz-main}. Its flat part matches the first term in Eq.~\eqref{eq:rhs-split-main}, the remaining part determines the curvature coefficient. The matching condition is
\begin{equation}
\label{eq:curv-matching-main}
-\zeta'(\lambda)\cV(\tau)G_\lambda^{(0)}(\tau)=H_{\rm curv}(\tau).
\end{equation}
Appendix~\ref{subsec:appendix-curv-flow} evaluates $H_{\rm curv}$ in the leading regular region and shows that the solution compatible with the initial condition \eqref{eq:zeta-initial-main} is
\begin{equation}
\label{eq:zeta-result-main}
\zeta(\lambda)=const =\frac{1}{12}.
\end{equation}
Thus the curvature coefficient already present in the first algebraic tail is preserved by the asymptotic link-exponential flow. At $\lambda=1$, Eqs.~\eqref{eq:flat-limit}, \eqref{eq:zeta-result-main} and using the flat normalization \eqref{eq:Gnorm-def} give
\begin{equation}
\label{eq:G_norm_curve}
G_{\rm norm}(a,b)
\simeq
\frac{1}{2}
-
\frac{1}{24}\cV(a,b)
+
\cdots
=
\frac{1}{2}
-
\frac{1}{24}
\int_{I(a,b)}\cR\,\d V
+
\cdots .
\end{equation}

\subsection{Identification of the effective curvature coupling}

The first-curvature correction in Eq.~\eqref{eq:G_norm_curve} has the same structure as the perturbative expansion of a continuum scalar Green's function with curvature coupling. Consider
\begin{equation}
(\Box_g-m^2-\xi\cR)\mathcal G(a,b)=\frac{\delta^{(2)}(a,b)}{\sqrt{-g}}.
\end{equation}
For $m=0$, expanding around the massless $1+1$-dimensional retarded Green's function gives
\begin{equation}
\mathcal G_\xi(a,b)
=
\mathcal G_0(a,b)
-
\xi\int_{I(a,b)}\d V\,\mathcal G_0(a,z)\,\cR\,\mathcal G_0(z,b)+\cdots .
\end{equation}
Since $\mathcal G_0=1/2$ inside the causal interval,
\begin{equation}
\mathcal G_\xi(a,b)
=
\frac{1}{2}
-
\frac{\xi}{4}\int_{I(a,b)}\cR\,\d V
+\cdots .
\end{equation}
Comparison with Eq.~\eqref{eq:G_norm_curve} gives
\begin{equation}
\xi_{\rm eff}=\frac{1}{6}.
\end{equation}
For a constant-curvature background, the curvature term can equivalently be absorbed into an
effective Klein--Gordon mass parameter,
\begin{equation}
m_{\mathrm{eff}}^2
\equiv
\xi_{\mathrm{eff}}\cR
=
\frac{\cR}{6}.
\label{eq:mind-main}
\end{equation}
This notation is useful for comparison with the massive continuum Green function, but
$m_{\mathrm{eff}}^2$ should not be interpreted as a physical particle mass. The primary result is
the effective curvature coupling $\xi_{\mathrm{eff}}=1/6$ generated by the link-exponential
kernel. Thus, to leading order in curvature, the normalized link-exponential kernel should be interpreted
as the retarded Green function associated with the effective operator
$\Box_g-\cR/6$, rather than with the minimally coupled massless operator $\Box_g$.
\begin{table}[t]
\centering
\renewcommand{\arraystretch}{1.35}
\setlength{\tabcolsep}{7pt}
\begin{tabular}{@{}lccc@{}}
\toprule
background & scalar curvature $\cR$ & effective mass $m_{\rm eff}^2=\cR/6$ & leading normalized kernel \\
\midrule
flat & $0$ & $0$ & $\displaystyle \frac{1}{2}$ \\[0.4em]
$\mathrm{AdS}_{1+1}$ & $\displaystyle -\frac{2}{\kappa^2}$ & $\displaystyle -\frac{1}{3\kappa^2}$ & $\displaystyle \frac{1}{2}+\frac{1}{6}\log\sec^2\!\left(\frac{\tau}{2\kappa}\right)$ \\[0.8em]
$\mathrm{dS}_{1+1}$ & $\displaystyle +\frac{2}{\kappa^2}$ & $\displaystyle +\frac{1}{3\kappa^2}$ & $\displaystyle \frac{1}{2}-\frac{1}{6}\log\cosh^2\!\left(\frac{\tau}{2\kappa}\right)$ \\
\bottomrule
\end{tabular}
\caption{Leading constant-curvature response of the normalized link-exponential kernel in flat, anti--de Sitter, and de Sitter backgrounds. The effective mass is fixed by $m_{\rm eff}^2=\xi_{\rm eff}\cR$ with $\xi_{\rm eff}=1/6$.}
\label{tab:ads-ds-summary}
\end{table}

The value $\xi_{\mathrm{eff}}=1/6$ is noteworthy for two independent reasons. First, it coincides numerically with the conformal curvature coupling of a scalar field in $3+1$ dimensions. This should not, however, be interpreted as a conformal-coupling result in the present $1+1$ dimensional setting, where the conformal value is
$\xi_{\mathrm{conf}}=0$.

Second, it is interesting from from the viewpoint of the local
short-distance expansion of scalar propagators~\cite{Vassilevich_2003}. For a massless scalar with curvature coupling $\xi R$, the first nontrivial local heat-kernel coefficient contains a curvature term proportional to $(1/6-\xi)\cR$. Hence the value found here is precisely the one for which this leading local scalar-curvature contribution vanishes. Since causal-set links are governed by the volumes of short Alexandrov intervals, this coincidence may indicate a connection with the ultraviolet geometric structure of continuum scalar propagation,
although establishing such a relation lies beyond the scope of the present work.

\section{Constant-curvature examples and numerical results}
\label{sec:constant-curvature-numerics}

The analytical calculation above fixes the leading coefficient. We now specialize to anti--de Sitter and de Sitter spacetime and compare the link-exponential data with the corresponding full continuum retarded solutions using the effective mass scale $m_{\rm eff}^2=\cR/6$.

\subsection{Anti--de Sitter and de Sitter examples}
\label{subsec:ads-ds}

Let $\kappa$ denote the curvature radius. In two spacetime dimensions,
\begin{equation}
\cR_{\rm AdS}=-\frac{2}{\kappa^2},
\qquad
\cR_{\rm dS}=+\frac{2}{\kappa^2}.
\end{equation}
Therefore Eq.~\eqref{eq:mind-main} gives
\begin{equation}
\label{eq:induced-masses-ads-ds}
m_{\rm eff,AdS}^2=-\frac{1}{3\kappa^2},
\qquad
m_{\rm eff,dS}^2=+\frac{1}{3\kappa^2}.
\end{equation}
Negative curvature enhances the kernel relative to the flat value, while positive curvature suppresses it.

For anti--de Sitter spacetime in $1+1$ dimensions we use conformal coordinates
\begin{equation}
ds^2=\frac{\kappa^2}{\cos^2 x}(-dt^2+dx^2).
\end{equation}
For a timelike interval of geodesic length $\tau$,
\begin{equation}
V_{\rm AdS}(\tau)=2\kappa^2\log\sec^2\left(\frac{\tau}{2\kappa}\right).
\end{equation}
Hence
\begin{equation}
\cV_{\rm AdS}(\tau)=\cR_{\rm AdS}V_{\rm AdS}(\tau)
=
-4\log\sec^2\left(\frac{\tau}{2\kappa}\right),
\end{equation}
and the leading order in curvature prediction becomes
\begin{equation}
G_{\rm norm}^{\rm AdS}(\tau)
\simeq
\frac{1}{2}
+
\frac{1}{6}\log\sec^2\left(\frac{\tau}{2\kappa}\right)+\cdots .
\end{equation}
For de Sitter spacetime in $1+1$ dimensions, the corresponding first-curvature prediction is
\begin{equation}
G_{\rm norm}^{\rm dS}(\tau)
\simeq
\frac{1}{2}
-
\frac{1}{6}\log\cosh^2\left(\frac{\tau}{2\kappa}\right)+\cdots .
\end{equation}
The sign difference and results for both spacetimes are summarized in Table~\ref{tab:ads-ds-summary}.
\begin{figure}[t!]
    \centering
    \subfigure[]{\includegraphics[width=0.495\textwidth]{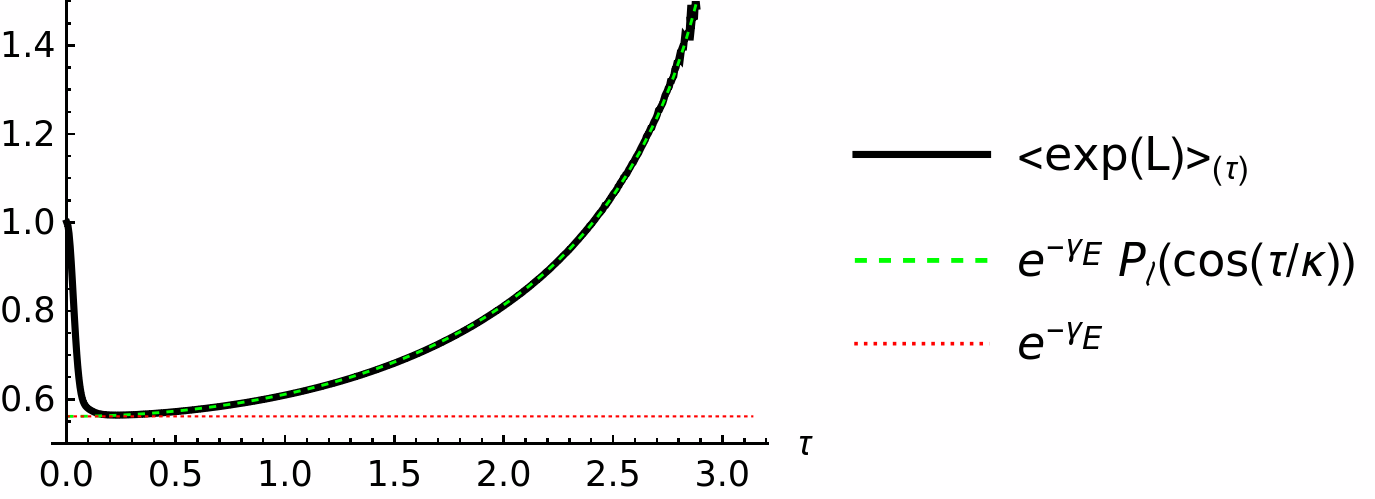}}
    \hspace{0.0001\textwidth}%
    \subfigure[]{\includegraphics[width=0.495\textwidth]{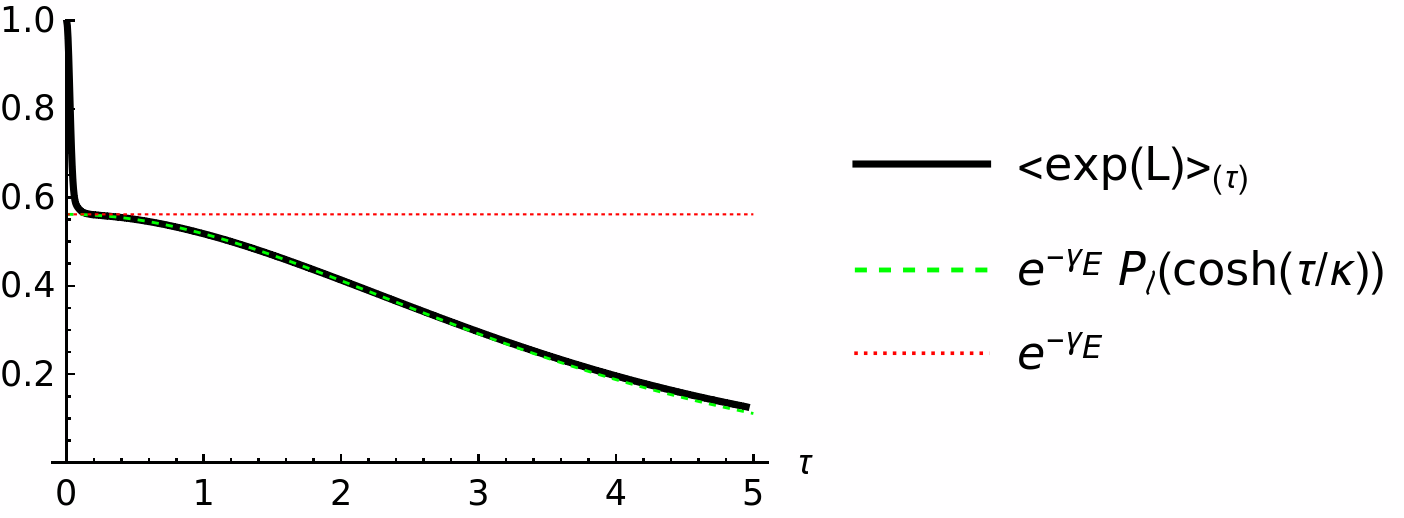}}%
    \caption{Numerical comparison of the link-exponential kernel $G(\tau)=\langle(e^{\LLL})_{ab}\rangle$ with retarded continuum solutions in constant-curvature backgrounds for $\kappa=1$, using fixed density $\rho=10^3$ and $10^4$ independent sprinkling averages. Panel (a) shows $\mathrm{AdS}_{1+1}$, where negative curvature gives $m_{\rm eff}^2<0$ contribution and enhances the propagator relative to the flat value. Panel (b) shows $\mathrm{dS}_{1+1}$, where positive curvature gives $m_{\rm eff}^2>0$ and suppresses the propagator. The continuum curves use the analytically predicted coupling $\xi_{\rm eff}=1/6$ and are rescaled to the normalization of the causal-set quantity. The red dotted line marks the flat asymptotic value $e^{-\gamma_E}$.}
    \label{fig:(A)dS_m=0}
\end{figure}
\subsection{Comparison with continuum retarded solutions}

For the numerical comparison we use the corresponding retarded continuum solutions with the effective coupling $\xi=\xi_{\rm eff}=1/6$. Writing $s=\tau/\kappa$, the radial equations inside the future light cone are
\begin{align}
\mathrm{AdS}_{1+1}:&\quad \mathcal G''(s)+\cot s\,\mathcal G'(s)-2\xi\,\mathcal G(s)=0,\\
\mathrm{dS}_{1+1}:&\quad \mathcal G''(s)+\coth s\,\mathcal G'(s)+2\xi\,\mathcal G(s)=0.
\end{align}
The retarded solutions normalized to $G^R(0)=1/2$ are
\begin{equation}
\mathcal G_{\xi}^{R,{\rm AdS}}(\tau)=\frac{1}{2}P_\nu\left(\cos\frac{\tau}{\kappa}\right),
\qquad
\mathcal G_{\xi}^{R,{\rm dS}}(\tau)=\frac{1}{2}P_\nu\left(\cosh\frac{\tau}{\kappa}\right),
\qquad
\nu(\nu+1)=-2\xi.
\end{equation}
For $\xi=1/6$, this gives $\nu(\nu+1)=-1/3$.
\begin{figure}[t!]
    \centering
    \subfigure[]{\includegraphics[width=0.49\textwidth]{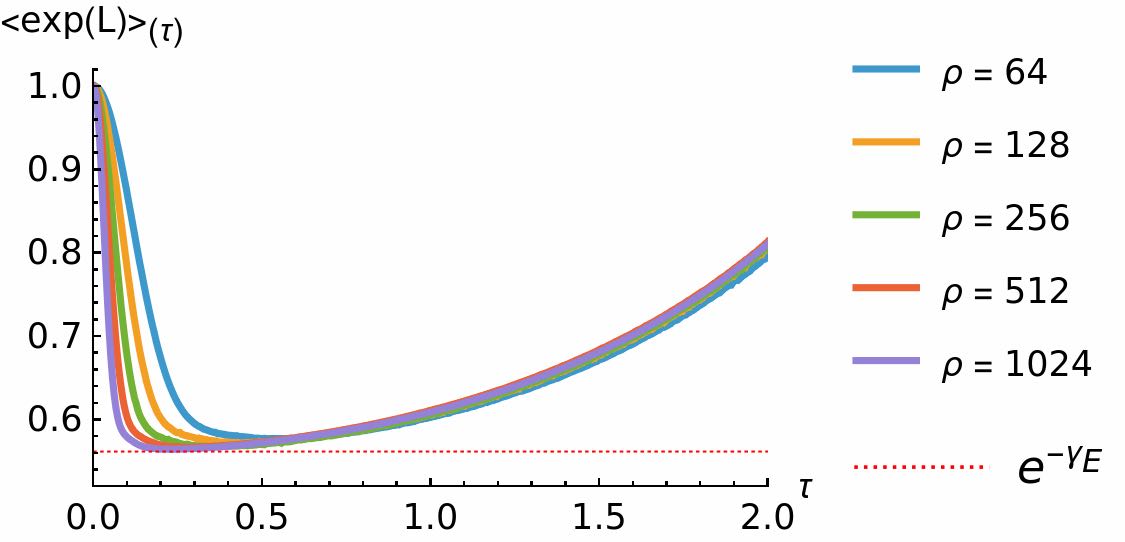}}
    \hspace{0.0001\textwidth}%
    \subfigure[]{\includegraphics[width=0.49\textwidth]{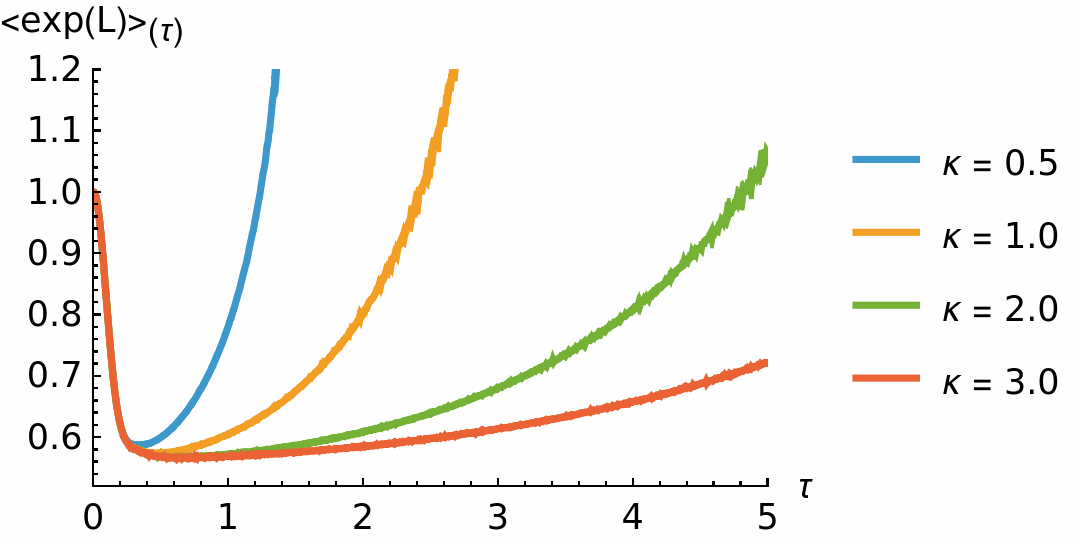}}%
   \caption{Density and curvature-scale dependence of the link-exponential kernel in $\mathrm{AdS}_{1+1}$ using $10^4$ independent sprinkling averages. Panel (a) shows the dependence on sprinkling density $\rho$ at fixed curvature radius. Increasing $\rho$ mainly changes the short-distance behavior, while the larger-$\tau$ curvature-enhanced tail remains stable. Panel (b) shows the dependence on the curvature radius $\kappa$ with fixed density $\rho=10^2$. Smaller $\kappa$ corresponds to larger $|\cR_{\rm AdS}|$ and hence to a larger magnitude of the effective scale $m_{\rm eff,AdS}^2=-1/(3\kappa^2)$.}
\label{fig:density_v_r}
\end{figure}

Fig.~\ref{fig:(A)dS_m=0} compares the causal-set quantity $G(\tau)=\langle(e^{\LLL})_{ab}\rangle$ with these continuum curves after rescaling by the inverse of the flat normalization. The mass parameter is not fitted; it is fixed by $m_{\rm eff}^2=\cR/6$. The comparison therefore tests whether the perturbatively derived coefficient continues to organize the full constant-curvature response over the simulated range.
Panel (a) shows the anti--de Sitter case. Since $\cR<0$, the effective effective mass squared is negative and the kernel is enhanced relative to the flat reference value. Panel (b) shows the de Sitter case. Since $\cR>0$, the effective effective mass squared is positive and the kernel is suppressed. In both cases the data follow the corresponding continuum retarded solution with the effective coupling.

\subsection{Density, curvature, and mass scaling}

Figure~\ref{fig:density_v_r} shows two consistency checks in $\mathrm{AdS}_{1+1}$. In panel (a), the curvature radius is kept fixed while the sprinkling density is varied. Increasing $\rho$ mainly changes the short-distance part of the kernel: for small $\tau$, the curves approach the asymptotic regime at different rates. At larger $\tau$, the curves approach the same curvature-enhanced tail within finite-density corrections. This suggests that the long-distance curvature response is not a finite-density artifact. Rather, it is generated by the link structure itself and survives the continuum limit as an effective curvature coupling. Panel (b) shows the dependence on the curvature radius $\kappa$. Smaller $\kappa$ corresponds to larger $|\cR_{\rm AdS}|$ and hence to a larger magnitude of the effective scale $m_{\rm eff,AdS}^2=-1/(3\kappa^2)$. Accordingly, the enhancement becomes stronger as the curvature radius decreases. Conversely, for larger $\kappa$ the curvature scale is weaker and the curve stays closer to the flat reference value over a longer range in $\tau$.

\begin{figure}[t!]
   {\includegraphics[width=0.55\textwidth]{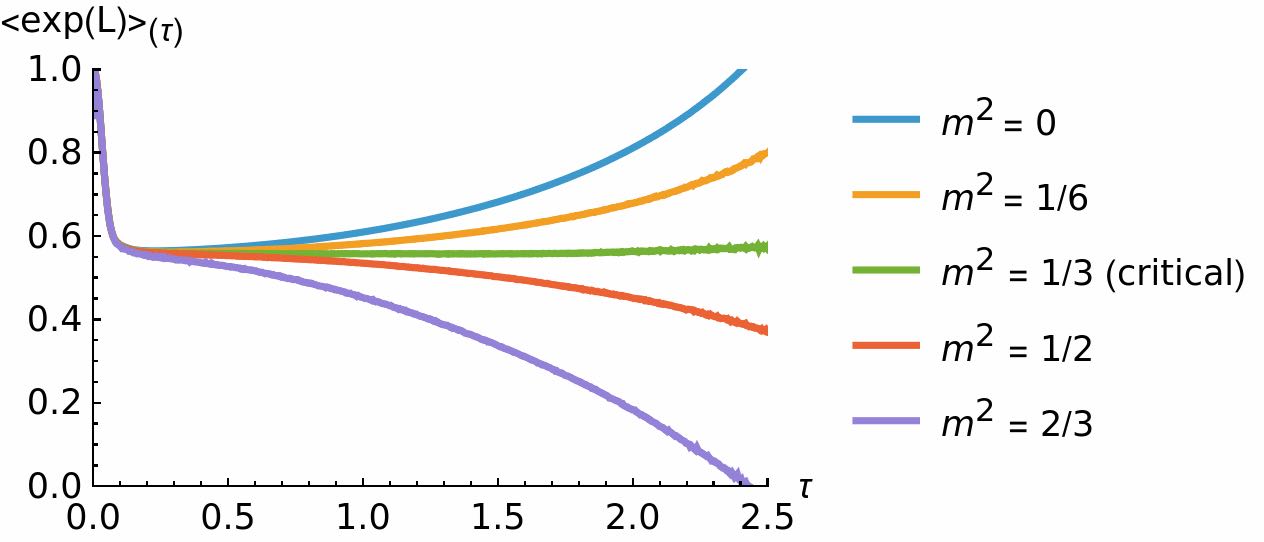}}%
  \caption{Mass dependence of the link-exponential kernel in $\mathrm{AdS}_{1+1}$ for $\kappa=1$ using fixed density $\rho=10^3$ and $10^4$ independent sprinkling averages. The curvature-induced contribution is $m_{\rm eff,AdS}^2=-1/3$. For $m^2=0$, the relative effective mass squared is negative and the propagator is enhanced. At $m^2=1/3$, the physical mass cancels the effective AdS contribution and the propagator becomes approximately flat. For $m^2>1/3$, the relative mass squared is positive and the propagator is suppressed.}
\label{fig:mass-dependence}
\end{figure}

The mass dependence, obtained by the usual mass scattering procedure, is shown in Fig.~\ref{fig:mass-dependence}. In AdS, the effective contribution is negative. A positive physical mass can therefore cancel it. For $\kappa=1$, we get $m_{\rm eff,AdS}^2=-1/3$. The cancellation occurs at $m^2=+1/3$, where the relative effective mass squared is approximately zero and the propagator becomes close to the flat massless behavior. For $m^2<1/3$, the remaining relative mass squared is negative and the kernel is enhanced. For $m^2>1/3$, it is positive and the kernel is suppressed. The observed mass dependence is consistent with the link-exponential propagator approaching the
continuum retarded Green function of the operator
\begin{equation}
\Box_g-m^2-\xi_{\rm eff}\cR, \quad \text{with}\quad\xi_{\rm eff}=1/6. 
\end{equation}
The curvature contribution is therefore already present in the link-exponential kernel itself, and the relative Klein--Gordon parameter is $m^2+\xi_{\rm eff}\cR$. Hence the explicit mass term does not act on a minimally coupled massless kernel, but on one that already carries the effective curvature contribution $\xi_{\rm eff}\cR$. The minimally coupled massless propagator is obtained when this contribution is cancelled, $m^2=-\xi_{\rm eff}\cR$.

\section{Conclusions}
\label{sec:conclusions}
We studied the response of the link-exponential causal-set propagator to curvature in $1+1$ dimensions. The construction is based on the link matrix $\LLL$, whose powers count paths built from irreducible causal relations. In flat spacetime, this construction was shown in Ref.~\cite{hinrichsen2026linkbasedcausalsetpropagators} to reproduce the massless retarded continuum propagator after the constant normalization $G_{\rm norm}=\frac12 e^{\gamma_E}G$. The aim of the present work was to test whether this agreement persists in curved conformally flat backgrounds.

The main mechanism is simple. In $1+1$ dimensions the conformal factor leaves the light cones
unchanged, so it does not affect causal accessibility. A link, however, requires the Alexandrov
interval between two causally related elements to be empty. Since the corresponding emptiness
probability is controlled by the physical interval volume, the link matrix remains sensitive to
curvature even though the minimally coupled massless continuum Green function is conformally
invariant.

Analytically, we derived the leading curvature correction
\begin{equation}
G_{\rm norm}(a,b)
\simeq
\frac{1}{2}
-
\frac{1}{24}\int_{I(a,b)}\cR\,\d V
+\cdots .\notag
\end{equation}
Comparison with a perturbative curvature-coupled continuum scalar gives
\begin{equation}
\xi_{\rm eff}=\frac{1}{6},
\qquad
m_{\rm eff}^2=\frac{\cR}{6}, \notag
\end{equation}
where the latter interpretation applies for constant curvature. Thus, to leading order in
curvature, the normalized link-exponential kernel is naturally associated with the retarded Green
function of the effective operator $\Box_g-\cR/6$, rather than with the minimally coupled
massless operator $\Box_g$.

This curvature coupling is generated by the interval-volume dependence of the link relation and
is not inserted by hand. The present
calculation alone does not establish whether the coefficient $1/6$ is specific to this construction
or reflects a more general property of link-based causal-set propagators.

An important feature is that the leading analytical curvature coefficient is independent of the
sprinkling density~$\rho$. The numerical density study is consistent with this result: increasing
$\rho$ primarily changes the short-distance crossover while leaving the large-distance curvature
response unchanged. Moreover, the AdS and dS simulations are
well described by the full constant-curvature continuum propagators with
$\xi=1/6$, suggesting that the identification with $\Box_g-\cR/6$ extends beyond the
perturbative regime in which we derived it analytically. Together, these results support the
interpretation that the curvature coupling is not a finite-density artifact but survives in the
high-density continuum limit.

In AdS spacetime the curvature is negative, so the effective mass squared is negative and the kernel is enhanced relative to the flat value. In dS spacetime the curvature is positive, so the effective mass squared is positive and the kernel is suppressed. The simulations support the stronger constant-curvature extrapolation: after fixing $m_{\rm eff}^2=\cR/6$ from the leading order calculation, the full continuum retarded solutions with this mass scale describe the data over the simulated range without fitting an additional mass parameter.

The analytical derivation remains pertubative and asymptotic. It derives the curvature coefficient in the regime $\rho\tau^2\gg1$ and $|\cR|\tau^2\ll1$. The numerical comparison suggests that this coefficient continues to control the finite-curvature constant-curvature curves, but this stronger behavior is an empirical observation within the simulated range. Curvature gradients, higher-curvature corrections, global effects, and possible curvature contributions from the singular part of the convolution kernel remain topics for further work.

The construction is related to, but distinct from, causal-set d'Alembertian and hop-and-stop propagator constructions~\cite{BenincasaDowker2010,DowkerGlaser2013,AslanbeigiSaravaniSorkin2014,Johnston2008,johnston2010quantumfieldscausalsets,XDowkerSurya2017,b4wn-qtvq}. A useful next step would be a systematic comparison of variance, convergence rate, and finite-density bias between link-exponential kernels and earlier causal-set propagators. It would also be interesting to investigate whether the inverse of the link-exponential kernel has an operator interpretation, as suggested in the flat analysis~\cite{hinrichsen2026linkbasedcausalsetpropagators}, and whether its curved corrections reproduce the effective curvature term found here. Finally, the resulting retarded kernel could be used as input for causal-set quantum field theory, where retarded propagators enter the construction of the Pauli--Jordan function and the Sorkin--Johnston two-point function~\cite{Johnston_2009,Afshordi_2012,PhysRevD.100.045007,Aslanbeigi_2013}.

A further question is whether the coefficient $\xi_{\mathrm{eff}}=1/6$
is peculiar to the present $1+1$ dimensional link-exponential kernel, or whether it reflects a
more general feature of propagators constructed directly from causal-set links. This question is
particularly motivated by existing higher-dimensional link-based constructions. In $3+1$
dimensions, the causal-set Green-function prescription of Nomaan X et. al. ~\cite{XDowkerSurya2017}
reproduces the conformally coupled massless scalar Green function in de Sitter spacetime and in a
conformally flat patch of anti--de Sitter spacetime, where the corresponding curvature coupling is
also $\xi=1/6$. Whether the recurrence of this value has a common origin in the microscopic
link structure, or is specific to the individual constructions and dimensions, remains an open
question.

\appendix
\section{Details of the asymptotic calculation}
\label{AppendixCalculation}

\subsection{Flat convolution and normalization}
\label{subsec:appendix-flat}

This subsection recalls the flat asymptotic estimate used in Sec.~\ref{sec:flat-reference}. It follows the analysis of Ref.~\cite{hinrichsen2026linkbasedcausalsetpropagators}, but the main steps are included here to make the present paper readable without consulting that work.

In flat spacetime the flow equation is
\begin{equation}
\partial_\lambda G_\lambda^{(0)}(\tau)=
(F_1^{(0)}\circledast G_\lambda^{(0)})(\tau),
\end{equation}
where $F_1^{(0)}(q)=e^{-\rho {\tau'}^2/2}$. In radial variables,
\begin{equation}
(F_1^{(0)}\circledast G_\lambda^{(0)})(\tau)
=
\int_0^\tau \d{\tau''}\int_0^{\tau-{\tau''}}\d{\tau'}\,K_0(\tau;{\tau'},{\tau''})F_1^{(0)}({\tau'})G_\lambda^{(0)}({\tau''}),
\end{equation}
with
\begin{equation}
K_0(\tau;{\tau'},{\tau''})=\frac{4\rho {\tau'}{\tau''}}{\sqrt{\Delta}},
\qquad
\Delta=[(\tau-{\tau''})^2-{\tau'}^2][(\tau+{\tau''})^2-{\tau'}^2].
\end{equation}
For the flat ansatz $G_\lambda^{(0)}(\tau)=\alpha(\lambda)\tau^{-\beta(\lambda)}$, the large-$\tau$ estimate gives
\begin{equation}
(F_1^{(0)}\circledast G_\lambda^{(0)})(\tau)
=
G_\lambda^{(0)}(\tau)(2\ln \tau+C_\lambda),
\end{equation}
where
\begin{equation}
C_\lambda
=
\ln\rho+
\ln2-
\gamma_E+
\psi\left(\frac{3-\beta(\lambda)}{2}\right)
-
\psi\left(\frac{2-\beta(\lambda)}{2}\right)
-
2\psi(2-\beta(\lambda)).
\end{equation}
Here $\psi$ is the digamma function. On the left-hand side,
\begin{equation}
\partial_\lambda G_\lambda^{(0)}(\tau)
=
\left(\frac{\alpha'}{\alpha}-\beta'\ln \tau\right)G_\lambda^{(0)}(\tau).
\end{equation}
Matching logarithmic terms gives $-\beta'=2$. With the small-$\lambda$ initial condition $\beta(0)=2$, this yields
\begin{equation}
\beta(\lambda)=2-2\lambda.
\end{equation}
The constant part gives $\alpha'/\alpha=C_\lambda$. Solving this differential equation gives
\begin{equation}
\alpha(\lambda)=
\frac{e^{-\lambda\gamma_E}(\rho/2)^{\lambda-1}}{\Gamma^2(\lambda)}.
\end{equation}
Hence $\beta(1)=0$ and $\alpha(1)=e^{-\gamma_E}$, which proves the flat asymptotic normalization in Eq.~\eqref{eq:flat-limit}.

\subsection{Small-diamond and null-corner calculation}
\label{subsec:appendix-small-diamond}

Write the local conformally flat metric as
\begin{equation}
\d s^2=e^{2\sigma(t,x)}(-\d t^2+\d x^2).
\end{equation}
The Ricci scalar is
\begin{equation}
\cR=2e^{-2\sigma}(\partial_t^2-\partial_x^2)\sigma .
\end{equation}
Choose local conformal coordinates centered at the midpoint $p$ of the interval, with
\begin{equation}
\sigma(p)=0,
\qquad
\partial_\mu\sigma(p)=0.
\end{equation}
A convenient local representative for constant curvature is
\begin{equation}
\sigma(t,x)=-\frac{\cR}{4}x^2+\orderof{x^3}.
\end{equation}
Indeed, $2(\partial_t^2-\partial_x^2)\sigma=\cR$ at the midpoint. Therefore
\begin{equation}
\Omega^2(t,x)=e^{2\sigma(t,x)}=1-\frac{\cR}{2}x^2+\cdots .
\end{equation}

Choose symmetric endpoints $a=(-\ell,0)$ and $b=(\ell,0)$ with $\ell=\tau/2$. Since the coordinates are conformal, the coordinate diamond is the flat diamond
\begin{equation}
D_0=\{(t,x): |x|\le \ell-|t|,\ -\ell\le t\le \ell\}.
\end{equation}
Hence
\begin{equation}
V(a,b)=\int_{D_0}\Omega^2(t,x)\,\d t\,\d x
=
\int_{D_0}\d t\,\d x-
\frac{\cR}{2}\int_{D_0}x^2\,\d t\,\d x+
\cdots .
\end{equation}
The first integral is $2\ell^2=\tau^2/2$, and the second is
\begin{equation}
\int_{D_0}x^2\,dt\,dx
=
\int_{-\ell}^{\ell}\left[\int_{-(\ell-|t|)}^{\ell-|t|}x^2\,\d x\right]\d t
=
\frac{\ell^4}{3}.
\end{equation}
Thus
\begin{equation}
V(a,b)=2\ell^2-\frac{\cR \ell^4}{6}+\cdots
=
\frac{\tau^2}{2}-\frac{\cR\tau^4}{96}+\cdots .
\end{equation}

We next compute the leading curvature correction to $F_2$. The two-link integral is Eq.~\eqref{eq:F2-def-main}. In null coordinates, the right null corner is
\begin{equation}
z_+:
\qquad
U=\ell,
\qquad
V=-\ell.
\end{equation}
A nearby point is written as
\begin{equation}
U=\ell-\alpha,
\qquad
V=-\ell+\beta,
\qquad
\alpha,\beta\ge0.
\end{equation}
Near this corner,
\begin{equation}
dV_z\simeq \frac12\Omega_+^2\,\d \alpha\,\d \beta,
\end{equation}
and the two small subinterval volumes are linear in the small widths,
\begin{equation}
V(a,z)\simeq A_+\beta,
\qquad
V(z,b)\simeq B_+\alpha,
\end{equation}
with
\begin{equation}
A_+=\frac12\int_{-\ell}^{\ell}\Omega^2(U,-\ell)\,\d U,
\qquad
B_+=\frac12\int_{-\ell}^{\ell}\Omega^2(\ell,V)\,\d V.
\end{equation}
Thus
\begin{equation}
F_{2,+}\simeq
\rho\frac{\Omega_+^2}{2}
\int_0^\infty \d\alpha\int_0^\infty \d\beta\,
e^{-\rho B_+\alpha}e^{-\rho A_+\beta}
=
\frac{\Omega_+^2}{2\rho A_+B_+}.
\end{equation}
Using $\Omega^2=1-\cR x^2/2+\cdots$, one finds
\begin{equation}
\Omega_+^2=1-\frac{\cR \ell^2}{2}+\cdots,
\qquad
A_+=B_+=\ell\left(1-\frac{\cR \ell^2}{6}\right)+\cdots .
\end{equation}
Therefore
\begin{equation}
F_{2,+}=\frac{1}{2\rho \ell^2}\left(1-\frac{\cR \ell^2}{6}+\cdots\right).
\end{equation}
The left corner gives the same leading contribution in constant curvature. Thus
\begin{equation}
F_2(a,b)=
\frac{1}{\rho \ell^2}\left(1-\frac{\cR \ell^2}{6}+\cdots\right)
=
\frac{4}{\rho\tau^2}\left(1-\frac{\cR\tau^2}{24}+\cdots\right).
\end{equation}

\subsection{Expansion of the curved convolution kernel}
\label{subsec:appendix-kernel}

In the curved convolution we use the radial variables of Eq.~\eqref{eq:radial-vars-main}, so that
\begin{equation}
\rho\int_{I(a,b)}\d V=\int d{\tau'}\,d{\tau''}\,K(\tau;{\tau'},{\tau''}).
\end{equation}
The flat kernel is
\begin{equation}
K_0(\tau;{\tau'},{\tau''})=\frac{4\rho {\tau'} {\tau''}}{\sqrt{\Delta}},
\qquad
\Delta=[(\tau-{\tau''})^2-{\tau'}^2][(\tau+{\tau''})^2-{\tau'}^2].
\end{equation}
In the local conformal gauge,
\begin{equation}
\Omega^2(t,X)=1-\frac{\cR}{2}x^2+\cdots .
\end{equation}
Choose the external points as $a=(-\tau/2,0)$ and $b=(\tau/2,0)$, and let $z=(t,X)$. Define the flat coordinate distances
\begin{equation}
{\tau'}^2_0=\left(t+\frac{\tau}{2}\right)^2-X^2,
\qquad
{\tau''}^2_0=\left(\frac{\tau}{2}-t\right)^2-X^2 .
\end{equation}
To first order in the conformal perturbation, the physical segment lengths satisfy
\begin{equation}
{\tau'}={\tau'}_0\left(1-\frac{\cR X^2}{12}+\cdots\right),
\qquad
{\tau''}={\tau''}_0\left(1-\frac{\cR X^2}{12}+\cdots\right).
\end{equation}
Equivalently,
\begin{equation}
{\tau'}_0={\tau'}\left(1+\frac{\cR X^2}{12}+\cdots\right),
\qquad
{\tau''}_0={\tau''}\left(1+\frac{\cR X^2}{12}+\cdots\right),
\end{equation}
and at this order
\begin{equation}
X^2=\frac{\Delta}{4\tau^2}.
\end{equation}
The curved kernel is obtained from
\begin{equation}
K(\tau;{\tau'},{\tau''})=\Omega^2(t,X)K_0(\tau;{\tau'}_0,{\tau''}_0)
\left|\frac{\partial({\tau'}_0,{\tau''}_0)}{\partial({\tau'},{\tau''})}\right|.
\end{equation}
Expanding the volume factor, the flat kernel, and the Jacobian gives
\begin{equation}
\frac{K(\tau;{\tau'},{\tau''})}{K_0(\tau;{\tau'},{\tau''})}
=
1+
\cR \left[
-\frac{X^2}{2}
+\frac{X^2}{6}
+\frac{1}{12}({\tau'}\partial_{{\tau'}}+{\tau''}\partial_{{\tau''}})X^2
+\frac{X^2}{12}({\tau'}\partial_{{\tau'}}+{\tau''}\partial_{{\tau''}})\ln K_0
\right]+
\cdots .
\end{equation}
Using $X^2=\Delta/(4\tau^2)$ and $K_0=4\rho {\tau'}{\tau''}/\sqrt{\Delta}$, the bracket evaluates to $-(\tau^2-{\tau'}^2-{\tau''}^2)/24$. Therefore
\begin{equation}
\label{eq:kernel-expansion-appendix}
K(\tau;{\tau'},{\tau''})=K_0(\tau;{\tau'},{\tau''})
\left[1-\frac{\cR}{24}(\tau^2-{\tau'}^2-{\tau''}^2)+\cdots\right].
\end{equation}
This gives the first term in Eq.~\eqref{eq:I-main-short}. The second term in Eq.~\eqref{eq:I-main-short} comes from the one-link correction
\begin{equation}
F_1({\tau'})=F_1^{(0)}({\tau'})\left[1+\frac{\rho\cR {\tau'}^4}{96}+\cdots\right].
\end{equation}
The third term comes from Eq.~\eqref{eq:smart-zero-main}. For constant curvature,
\begin{equation}
\zeta[\cV(\tau)-\cV({\tau''})]=\zeta\frac{\cR}{2}(\tau^2-{\tau''}^2).
\end{equation}

\subsection{Curvature-flow matching}
\label{subsec:appendix-curv-flow}

We now evaluate the matching condition \eqref{eq:curv-matching-main}. The link factor localizes the first segment at
\begin{equation}
{\tau'}\sim \rho^{-1/2}.
\end{equation}
Therefore explicit powers of $q$ in Eq.~\eqref{eq:I-main-short} are suppressed by $1/(\rho \tau^2)$ relative to the leading curvature scale $\cR_0T^2$. Keeping the leading regular-region source gives
\begin{equation}
\cI_{\rm lead}(\tau;{\tau''})=
\cR(\tau^2-{\tau''}^2)\left(\frac{\zeta}{2}-\frac{1}{24}\right).
\end{equation}
In the regular region $q\ll |\tau-{\tau''}|$ and ${\tau'}\ll \tau+{\tau''}$, the flat kernel reduces to
\begin{equation}
K_0(\tau;{\tau'},{\tau''})\simeq \frac{4\rho {\tau'}{\tau''}}{\tau^2-{\tau''}^2}.
\end{equation}
After integrating over the leading one-link factor,
\begin{equation}
\widetilde K_0(\tau,{\tau''})
=
\int_0^\infty \d{\tau'}\,K_0(\tau;{\tau'},{\tau''})e^{-\rho {\tau''}^2/2}
\simeq
\frac{4{\tau''}}{\tau^2-{{\tau''}}^2}.
\end{equation}
The curvature source is then
\begin{equation}
H_{\rm curv}(\tau)
=
\int_0^\tau \d{\tau''}\,\widetilde K_0(\tau,{\tau''})G_\lambda^{(0)}({\tau''})\cI_{\rm lead}(\tau;{\tau''}).
\end{equation}
The factor $\tau^2-{\tau''}^2$ cancels, so
\begin{equation}
H_{\rm curv}(\tau)=
\cR\left(2\zeta-\frac{1}{6}\right)
\int_0^\tau {\tau''}G_\lambda^{(0)}({\tau''})\,\d{\tau''}.
\end{equation}
With $G_\lambda^{(0)}({\tau''})=\alpha {\tau''}^{-\beta}$ and $\beta(\lambda)=2-2\lambda$,
\begin{equation}
\int_0^\tau {\tau''}G_\lambda^{(0)}({\tau''})\,d{\tau''}
=
\frac{\tau^2G_\lambda^{(0)}(\tau)}{2\lambda}.
\end{equation}
Thus
\begin{equation}
H_{\rm curv}(\tau)=
\cR\left(2\zeta-\frac{1}{6}\right)
\frac{\tau^2G_\lambda^{(0)}(\tau)}{2\lambda}.
\end{equation}
On the left-hand side of Eq.~\eqref{eq:curv-matching-main}, constant curvature gives
\begin{equation}
\cV(\tau)=\frac{\cR \tau^2}{2}.
\end{equation}
Substitution into the matching condition and cancellation of the common factor $\cR \tau^2G_\lambda^{(0)}(\tau)$ gives
\begin{equation}
\zeta'(\lambda)=\frac{2}{\lambda}\left(\frac{1}{12}-\zeta(\lambda)\right).
\end{equation}
Equivalently,
\begin{equation}
\zeta'(\lambda)+\frac{2}{\lambda}\zeta(\lambda)=\frac{1}{6\lambda}.
\end{equation}
The integrating factor is $\lambda^2$, so
\begin{equation}
(\lambda^2\zeta)'=\frac{1}{6}\lambda.
\end{equation}
Thus
\begin{equation}
\zeta(\lambda)=\frac{1}{12}+\frac{C}{\lambda^2}.
\end{equation}
Regularity as $\lambda\to0^+$, together with $\zeta(0^+)=1/12$, gives $C=0$. Hence $\zeta(\lambda)=1/12$ on the asymptotic branch.

The regular-region approximation used here does not analyze possible curvature sources from the singular region $T-r\sim q$. In flat spacetime that region is responsible for the logarithmic contribution in Eq.~\eqref{eq:flat-conv-result-main}. The present calculation extracts the leading curvature source; a full treatment of singular-region curvature corrections remains an open technical point.


\begin{acknowledgments}
This work was supported financially by the Employment Agency Bonn and the University of Würzburg.
The author thanks H. Hinrichsen for critical reading of the manuscript.
\end{acknowledgments}

\bibliography{references}

@book{surya2025causal,
  title={The Causal Set Approach to Quantum Gravity: An Introduction},
  author={Surya, Sumati},
  year={2025},
  publisher={Springer Nature}
}

@article{Bombelli1987,
  author  = {Bombelli, Luca and Lee, Joohan and Meyer, David and Sorkin, Rafael D.},
  title   = {Space-Time as a Causal Set},
  journal = {Physical Review Letters},
  volume  = {59},
  number  = {5},
  pages   = {521--524},
  year    = {1987},
  doi     = {10.1103/PhysRevLett.59.521}
}

@misc{Sorkin2003,
  author = {Sorkin, Rafael D.},
  title  = {Causal Sets: Discrete Gravity},
  year   = {2003},
  eprint = {gr-qc/0309009},
  archivePrefix = {arXiv}
}

@article{Surya2019,
  author  = {Surya, Sumati},
  title   = {The Causal Set Approach to Quantum Gravity},
  journal = {Living Reviews in Relativity},
  volume  = {22},
  number  = {1},
  year    = {2019},
  doi     = {10.1007/s41114-019-0023-1}
}

@article{Malament1977,
  author  = {Malament, David B.},
  title   = {The Class of Continuous Timelike Curves Determines the Topology of Spacetime},
  journal = {Journal of Mathematical Physics},
  volume  = {18},
  number  = {7},
  pages   = {1399--1404},
  year    = {1977},
  doi     = {10.1063/1.523436}
}

@article{HawkingKingMcCarthy1976,
  author  = {Hawking, S. W. and King, A. R. and McCarthy, P. J.},
  title   = {A New Topology for Curved Space-Time Which Incorporates the Causal, Differential, and Conformal Structures},
  journal = {Journal of Mathematical Physics},
  volume  = {17},
  number  = {2},
  pages   = {174--181},
  year    = {1976},
  doi     = {10.1063/1.522874}
}

@book{BirrellDavies1982,
  author    = {Birrell, N. D. and Davies, P. C. W.},
  title     = {Quantum Fields in Curved Space},
  publisher = {Cambridge University Press},
  year      = {1982}
}

@book{Wald1994,
  author    = {Wald, Robert M.},
  title     = {Quantum Field Theory in Curved Spacetime and Black Hole Thermodynamics},
  publisher = {University of Chicago Press},
  year      = {1994}
}

@article{Johnston2008,
  author  = {Johnston, Steven},
  title   = {Particle Propagators on Discrete Spacetime},
  journal = {Classical and Quantum Gravity},
  volume  = {25},
  number  = {20},
  pages   = {202001},
  year    = {2008},
  doi     = {10.1088/0264-9381/25/20/202001}
}

@article{BenincasaDowker2010,
  author  = {Benincasa, Dionigi M. T. and Dowker, Fay},
  title   = {Scalar Curvature of a Causal Set},
  journal = {Physical Review Letters},
  volume  = {104},
  number  = {18},
  pages   = {181301},
  year    = {2010},
  doi     = {10.1103/PhysRevLett.104.181301}
}

@article{DowkerGlaser2013,
  author  = {Dowker, Fay and Glaser, Lisa},
  title   = {Causal Set d'Alembertians for Various Dimensions},
  journal = {Classical and Quantum Gravity},
  volume  = {30},
  number  = {19},
  pages   = {195016},
  year    = {2013},
  doi     = {10.1088/0264-9381/30/19/195016}
}

@article{AslanbeigiSaravaniSorkin2014,
  author  = {Aslanbeigi, Siavash and Saravani, Mehdi and Sorkin, Rafael D.},
  title   = {Generalized Causal Set d'Alembertians},
  journal = {Journal of High Energy Physics},
  volume  = {2014},
  number  = {6},
  year    = {2014},
  doi     = {10.1007/JHEP06(2014)024}
}

@article{XDowkerSurya2017,
  title={Scalar field Green functions on causal sets},
   volume={34},
   ISSN={1361-6382},
   url={http://dx.doi.org/10.1088/1361-6382/aa6bc7},
   DOI={10.1088/1361-6382/aa6bc7},
   number={12},
   journal={Classical and Quantum Gravity},
   publisher={IOP Publishing},
   author={X, Nomaan and Dowker, Fay and Surya, Sumati},
   year={2017},
   month=May, pages={124002} }

@article{AvisIshamStorey1978,
  author  = {Avis, S. J. and Isham, C. J. and Storey, D.},
  title   = {Quantum Field Theory in Anti-de Sitter Space-Time},
  journal = {Physical Review D},
  volume  = {18},
  number  = {10},
  pages   = {3565--3576},
  year    = {1978},
  doi     = {10.1103/PhysRevD.18.3565}
}

@article{BreitenlohnerFreedman1982,
  author  = {Breitenlohner, Peter and Freedman, Daniel Z.},
  title   = {Stability in Gauged Extended Supergravity},
  journal = {Annals of Physics},
  volume  = {144},
  number  = {2},
  pages   = {249--281},
  year    = {1982},
  doi     = {10.1016/0003-4916(82)90116-6}
}

@article{Maldacena1998,
  author  = {Maldacena, Juan M.},
  title   = {The Large N Limit of Superconformal Field Theories and Supergravity},
  journal = {Advances in Theoretical and Mathematical Physics},
  volume  = {2},
  pages   = {231--252},
  year    = {1998},
  eprint  = {hep-th/9711200},
  archivePrefix = {arXiv}
}

@article{Witten1998,
  author  = {Witten, Edward},
  title   = {Anti de Sitter Space and Holography},
  journal = {Advances in Theoretical and Mathematical Physics},
  volume  = {2},
  pages   = {253--291},
  year    = {1998},
  eprint  = {hep-th/9802150},
  archivePrefix = {arXiv}
}

@article{Aharony2000,
  author  = {Aharony, Ofer and Gubser, Steven S. and Maldacena, Juan M. and Ooguri, Hirosi and Oz, Yaron},
  title   = {Large N Field Theories, String Theory and Gravity},
  journal = {Physics Reports},
  volume  = {323},
  number  = {3--4},
  pages   = {183--386},
  year    = {2000},
  doi     = {10.1016/S0370-1573(99)00083-6},
  eprint  = {hep-th/9905111},
  archivePrefix = {arXiv}
}

@misc{hinrichsen2026linkbasedcausalsetpropagators,
      title={Link-based causal set propagators in $1+1$ dimensions}, 
      author={Haye Hinrichsen and Arsim Kastrati},
      year={2026},
      eprint={2604.24812},
      archivePrefix={arXiv},
      primaryClass={gr-qc},
      url={https://arxiv.org/abs/2604.24812}, 
}

@misc{johnston2010quantumfieldscausalsets,
      title={Quantum Fields on Causal Sets}, 
      author={Steven Johnston},
      year={2010},
      eprint={1010.5514},
      archivePrefix={arXiv},
      primaryClass={hep-th},
      url={https://arxiv.org/abs/1010.5514}, 
}

@article{b4wn-qtvq,
  title = {Numerical evaluation of the causal set propagator in $2D$ anti--de Sitter spacetime},
  author = {Kastrati, Arsim and Hinrichsen, Haye},
  journal = {Phys. Rev. D},
  volume = {113},
  issue = {6},
  pages = {065003},
  numpages = {13},
  year = {2026},
  month = {Mar},
  publisher = {American Physical Society},
  doi = {10.1103/b4wn-qtvq},
  url = {https://link.aps.org/doi/10.1103/b4wn-qtvq}
}

@article{Johnston_2009,
   title={Feynman Propagator for a Free Scalar Field on a Causal Set},
   volume={103},
   ISSN={1079-7114},
   url={http://dx.doi.org/10.1103/PhysRevLett.103.180401},
   DOI={10.1103/physrevlett.103.180401},
   number={18},
   journal={Physical Review Letters},
   publisher={American Physical Society (APS)},
   author={Johnston, Steven},
   year={2009},
   month=Oct }

@article{Afshordi_2012,
   title={A distinguished vacuum state for a quantum field in a curved spacetime: formalism, features, and cosmology},
   volume={2012},
   ISSN={1029-8479},
   url={http://dx.doi.org/10.1007/JHEP08(2012)137},
   DOI={10.1007/jhep08(2012)137},
   number={8},
   journal={Journal of High Energy Physics},
   publisher={Springer Science and Business Media LLC},
   author={Afshordi, Niayesh and Aslanbeigi, Siavash and Sorkin, Rafael D.},
   year={2012},
   month=Aug }

@article{PhysRevD.100.045007,
  title = {Sorkin-Johnston vacuum for a massive scalar field in the 2D causal diamond},
  author = {Mathur, Abhishek and Surya, Sumati},
  journal = {Phys. Rev. D},
  volume = {100},
  issue = {4},
  pages = {045007},
  numpages = {31},
  year = {2019},
  month = {Aug},
  publisher = {American Physical Society},
  doi = {10.1103/PhysRevD.100.045007},
  url = {https://link.aps.org/doi/10.1103/PhysRevD.100.045007}
}

@article{Aslanbeigi_2013,
   title={A preferred ground state for the scalar field in de Sitter space},
   volume={2013},
   ISSN={1029-8479},
   url={http://dx.doi.org/10.1007/JHEP08(2013)039},
   DOI={10.1007/jhep08(2013)039},
   number={8},
   journal={Journal of High Energy Physics},
   publisher={Springer Science and Business Media LLC},
   author={Aslanbeigi, S. and Buck, M.},
   year={2013},
   month=Aug }

@article{Vassilevich_2003,
   title={Heat kernel expansion: user’s manual},
   volume={388},
   ISSN={0370-1573},
   url={http://dx.doi.org/10.1016/j.physrep.2003.09.002},
   DOI={10.1016/j.physrep.2003.09.002},
   number={5-6},
   journal={Physics Reports},
   publisher={Elsevier BV},
   author={Vassilevich, D.V.},
   year={2003},
   month=Dec, pages={279–360} }

\end{document}